\documentclass[manuscript, screen]{acmart}

\usepackage{tabularx}
\usepackage{graphicx}
\usepackage{caption}

\copyrightyear{2026}
\acmYear{2026}
\setcopyright{cc}
\setcctype{by}
\acmConference[CHI EA '26]{Extended Abstracts of the 2026 CHI Conference on Human Factors in Computing Systems}{April 13--17, 2026}{Barcelona, Spain}
\acmBooktitle{Extended Abstracts of the 2026 CHI Conference on Human Factors in Computing Systems (CHI EA '26), April 13--17, 2026, Barcelona, Spain}
\acmDOI{10.1145/3772363.3798719}
\acmISBN{979-8-4007-2281-3/2026/04}

\begin{document}

\title{Shifting Engagement With Cybersecurity: How People Discover and Share Cybersecurity Content at Work and at Home}
\renewcommand{\shorttitle}{Shifting Engagement with Cybersecurity}

\author{William Seymour}
\email{william.seymour@kcl.ac.uk}
\orcid{0000-0002-0256-6740}
\affiliation{%
  \institution{King's College London}
  \city{London}
  \country{UK}
}

\author{Martin J. Kraemer}
\email{mkraemer@knowbe4.com}
\orcid{0000-0001-7115-1066}
\affiliation{%
  \institution{KnowBe4}
  \city{London}
  \country{UK}
}

\renewcommand{\shortauthors}{Seymour and Kraemer}

\ccsdesc[500]{Security and privacy~Social aspects of security and privacy}

\keywords{Cybersecurity, News, Sharing, Workplace Training, Phishing, Social Engineering}

\begin{abstract}
Cybersecurity awareness is shaped by a wide range of professional and personal experiences, including information and training at work and the sharing of news and other content at home. In order to explore how people discover cybersecurity content and the effect that participation in workplace training may have on this we present an online study of 1200 participants from the UK, US, France, and Germany. Those undertaking cybersecurity training at work showed reduced intention to share information at home, shifting the focus towards the workplace. They were also more likely to recall cybersecurity information shared by their employer than from any other source, which in turn correlated with content type and distribution channel. We critically reflect on this shift, highlighting opportunities to improve cybersecurity information sharing at work and at home.
\end{abstract}

\maketitle

\section{Introduction \& Background}
While technical security measures generally evolve to become increasingly invisible and automatic (e.g. using disk encryption by default), people regularly and routinely deviate from ideal security and privacy behaviours for a variety of reasons. It is therefore not surprising that social engineering is ``a preferred attack vector for threat actors due to its simplicity, low cost, ease of execution, and profitability''~\cite{enisa2023}. Unlike other more specialised attack vectors, social engineering is a risk to \textit{everyone}, spanning governments, organisations, and individuals. Industry data suggests that the `human factor' contributes to over 60\% of all data breaches~\cite{verizon2024} and represents the initial access vector in 22\% of global cyber attacks (40\% in EMEA)~\cite{mandiant2023}). Social engineering poses a unique challenge by not being as easily mitigated through technical controls, making the \textit{information} people have a deciding factor in how they respond to threats~\cite{enisa2023}.

People encounter cybersecurity information from a variety of sources at home and at work, commonly sharing what they find with others~\cite{10.1145/3173574.3173575}. Looking closer, personal preferences, age, and life situation shape people's security and privacy information diets and motivations for sharing information, as well as the broad range of security and privacy media that people consume~\cite{10.1145/3173574.3173575}. Personal preferences include value-driven considerations, for example someone might value information that is readily available over information that is more authoritative but difficult to access~\cite{nicholson2019if}. This has previously been studied in exposure to news, in which cybersecurity is increasingly represented~\cite{das2014effect, 10.1145/3173574.3173575}, but the varying quality and comprehensiveness of online content means that workplace training and public awareness campaigns are also vital to support people in recognising threats.

The security awareness training (SAT) industry teaches employees the signs and tactics used in social engineering attacks against businesses and is projected to be worth \$10B by 2027~\cite{cybersecurityventures2023}. In recent years the importance of security awareness in the workplace has been reinforced by the GDPR; inclusion of employee training in NIS2 and DORA in the EU;\footnote{NIS2 is the updated and expanded EU Network and Information Security Directive and DORA is the EU Digital Operational Resilience Act.} and shift to remote working from the Covid-19 pandemic. While there is a temptation for businesses and training providers to consider cybersecurity only as a workplace issue, previous studies have demonstrated the often porous nature of the work-life boundary with people often reusing passwords and devices between work and personal matters~\cite{10.1145/2702123.2702386, 10.1145/3313831.3376881}. Practices such as account and device sharing between people are also commonplace~\cite{10.1145/2858036.2858051, 10.1145/3613904.3642874} and mean that personal, work, family, and community security postures are easily intertwined.

This suggests people might have a more complex cybersecurity `information diet' than previously thought. To explore this broader consumption of cybersecurity content we present the results of an exploratory survey of 1200 participants across the US, France, Germany, and the UK capturing a snapshot of the content participants had encountered, how they came across it, and whether they shared it onwards in order to answer the following research questions:

\begin{enumerate}
    \item[RQ1] How do people discover and share cybersecurity content?
    \item[RQ2] What is the effect of the workplace and SAT on the cybersecurity content people consume and share?
    \end{enumerate}

Our analysis extends our understanding of how people discover and share cybersecurity content and considers the impact of the workplace for the first time, finding that employers were the most common source of cybersecurity information and showing how different types of information are discovered and shared. We explore the correlation between participation in workplace cybersecurity training and increased information sharing and consumption within the workplace and make recommendations for employers, training providers, and avenues for future work.

\section{Methodology}
\subsection{Survey Design \& Recruitment}
To answer the research questions we designed a survey to understand how people discover and share cybersecurity content. The survey opened by asking participants to recall the most recent piece of cybersecurity content they remembered consuming. For this we used questions from Das et. al.~\cite{10.1145/3173574.3173575} on what the content was, how they had discovered it, who had shared it with them and why, and if they had shared it with anyone else (questions 4--10 \& 12 in~\cite{10.1145/3173574.3173575}). We also asked participants if they had encountered cybersecurity content through training at work, including factual questions about the type, frequency, and topics of that training. The survey closed with questions about participants' general privacy and security practices and attitudes. These were taken from the Internet Users’ Information Privacy Concerns inventory (IUIPC)~\cite{malhotra2004internet} measuring attitudes towards control, awareness, and data collection in the context of online privacy, and the SA-6 inventory of security attitudes~\cite{faklaris2019self}. IUIPC was developed and validated based on data from 742 interviews, and SA-6 on 687 survey responses. The survey and analysis are included as supplemental material.

A pilot study of 50 participants yielded usability and data quality improvements, and translations were written and checked by colleagues fluent in the target languages. The survey was administered using Qualtrics\footnote{\url{www.qualtrics.com}} with 1200 participants recruited using Prolific,\footnote{\url{www.prolific.com}} equally distributed between residents of the US, France, Germany, and the UK. We screened participants to ensure that they were fluent in the language the survey was written in and reported using a digital device at least weekly. The survey received ethical approval from the first author's institution, with participants paid above the `real living wage' in the first author's country given the mean completion time of 7m55s.

\subsection{Analysis}
Analysis and visualization were performed using Python notebooks and RStudio. To control for demographic factors we used multinomial log-linear regression models, included as Tables~\ref{tab:regression-how-inward-shared}, \ref{tab:regression-howdiscovered}, and \ref{tab:regression-whether-onward-shared} in the Appendix. Chi-squared tests were used where appropriate to test the independence of categorical responses to the survey questions. For IUIPC and SA-6 scores we calculated point-biserial coefficients to determine their correlation with participant behaviours (e.g., sharing a piece of content). Threat types were coded from participant content descriptions: the authors worked together to create a codebook from a sample of 200 responses before coding the rest independently according to language proficiency. Most responses were trivial to code with the remainder discussed and resolved during regular meetings. We took several measures to ensure the robustness of the results, including attention checks, filtering responses where participants did not recall cybersecurity content, and applying the Benjamini–Hochberg adjustment to regression models to correct for making multiple comparisons~\cite{benjamini1995controlling}.

\subsection{Limitations}\label{sec:limitations}
Whilst we prompted participants about potential types of content they may have consumed to increase recall and communicate what was in scope, this may have increased the likelihood of them recalling those items in the prompts. The alternative would have been to control the content participants were asked about (as in~\cite{10.1145/3173574.3173575}), but this would have ruled out studying workplace training. Alternative methods such as experience sampling~\cite{keusch2022using} would have reduced participation rates and required significantly increased resources, requiring scaling back the size of the participant pool and multi-lingual/national nature of the study which was undesirable given the noted US-centricity of research in this space~\cite{10.1145/3411764.3445488}. The snapshot nature of the study also meant that it was not possible to know how frequently individual participants shared information, although we expect the size of the sample to balance this.

\section{Results}

\newcommand\zz[1]{\multicolumn{#1}{>{\hsize=%
\dimexpr\numexpr#1\relax\hsize+
\numexpr#1*2-2\relax\tabcolsep+
\numexpr#1-1\relax\arrayrulewidth
\relax}X}}

After excluding low quality answers, the survey had 1095 respondents with an average age across the survey of 36.6 (Table~\ref{tab:age-demographics}). The majority of respondents reported using internet-connected devices multiple times every day (61\%), followed by every day (31\%), 2-6 times a week (8\%), and once every week (>1\%). Most were in full-time employment (57\%), followed by part-time employment (18\%), job seeking (7\%) or not in paid work (7\%), and 11\% stated other or undisclosed. Scores for the IUIPC and SA6 inventories showed an even distribution of security concerns and a strong skew towards privacy concerns.

\begin{figure*}
\centering
\begin{minipage}{.5\textwidth}
    \centering
    \begin{tabular}{l|l|l|l|l}
        Country & $n$ & $\bar{age}$ & $\sigma$ age & \% women/men \\
        \hline
        All & 1095 & 36.6 & 12.4 & 49.9/50.1 \\
        France & 277 & 30.9 & 9.9 & 50.5/49.5 \\
        Germany & 238 & 32.5 & 10.6 & 50.0/50.0 \\
        UK & 288 & 40.6 & 12.2 & 49.0/51.0 \\
        US & 292 & 41.3 & 12.9 & 50.0/50.0 \\
    \end{tabular}
    \captionof{table}{Participant age and gender identity by country}
    \label{tab:age-demographics}
\end{minipage}%
\begin{minipage}{.5\textwidth}
\begin{tabular}{l|r|r}
        Source & \% (rank) & \cite{10.1145/3173574.3173575} \% (rank) \\
        \hline
        Employers & 22\% (1) & - (-) \\
        Websites & 21\% (2) & 70\% (1) \\
        Social Media & 15\% (=3) & 29\% (3) \\
        Direct Sharing & 15\% (=3) & 17\% (4) \\
        Broadcasts/Podcasts & 13\% (5) & 36\% (2) \\
        Companies & 8\% (6) & 4\% (=5) \\
        Other & 6\% (7) & 4\% (=5) \\
    \end{tabular}
    \captionof{table}{Comparison of content source between the present study and \cite{10.1145/3173574.3173575}. Note that figures from \cite{10.1145/3173574.3173575} sum to over 100\%.}
    \label{tab:disovery_comparison}
\end{minipage}%
\end{figure*} 

\subsection{How People Discover and Share Cybersecurity Content (RQ1)}
\subsubsection{Content Discovery.}
\begin{figure}
\centering
\begin{minipage}{.5\textwidth}
    \centering
    \includegraphics[width=0.8\textwidth]{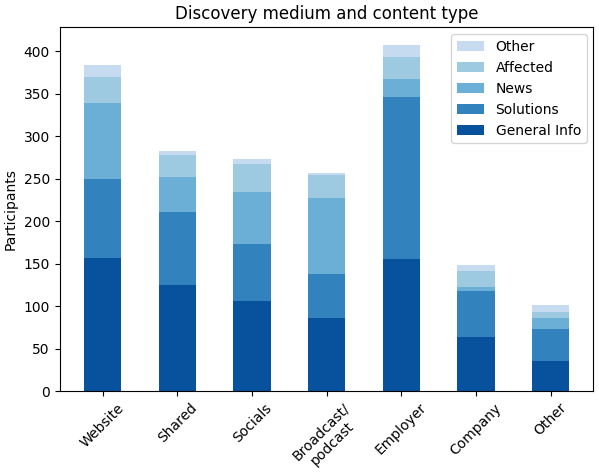}
    \caption{How participants discovered content.}
    \label{fig:discovery-medium}
\end{minipage}%
\begin{minipage}{.5\textwidth}
 \centering
    \includegraphics[width=\textwidth]{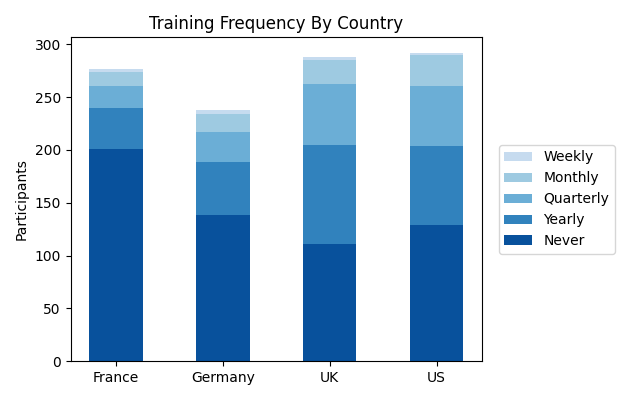}
    \caption{Engagement with workplace cybersecurity training.}
    \label{fig:workplace-training-frequency}
\end{minipage}%
\end{figure}

We began by exploring the relationship between the content that participants discovered and how they found it.  When asked about the last piece of cybersecurity content they had consumed, participants had mostly found this through their employers (22.0\%) or websites (21.1\%), followed by direct sharing (15.0\%, see below), social media (14.5\%), and broadcasts/podcasts (13.2\%), with messaging from other companies (8.3\%) less common (Figure~\ref{fig:discovery-medium}). We found that employers were more likely than any other source to share what were perceived as `solutions', and that broadcasts and podcasts were the most likely to communicate news.

Comparing these results with those of Das et. al.~\cite{10.1145/3173574.3173575} shows a considerable difference in the way that cybersecurity news is discovered and shared (Table \ref{tab:disovery_comparison}).\footnote{When comparing it is worth noting that we broaden the object of study from \textit{news} to \textit{content} and change the framing from \textit{security \& privacy} to \textit{cybersecurity}, in both cases to reflect the evolving zeitgeist. \textit{Television news} was broadened to \textit{broadcasts and podcasts}. Das et. al.~\cite{10.1145/3173574.3173575} did not consider employers or podcasts as sources of news, although this is instructive in itself as to their rise in importance---industry statistics suggest the cybersecurity awareness training market and podcast listening have roughly doubled since 2018~\cite{cybersecurityventures2023, edison2024}.} In the present study, employers have joined websites as the most prevalent source of discovery and direct sharing and sharing by companies was higher. To balance this, discovery via broadcasts was much lower, even when including podcasts and radio in addition to television. The proportion of people receiving content directly from others is similar, but we found much higher rates of sharing from colleagues (21\% vs 13\%) balanced by lower sharing from friends (32\% from 64\%). Messaging replaced face-to-face communication as the most common source of direct sharing. Focussing on the specific threat covered by the content, participants' most common combinations of source of information and threat type were learning about phishing from their employer (7.5\%); reading about data breaches on websites (5.5\%); generally recalling awareness training at work without further specification of content (5.3\%); having learned about data breaches from TV, radio, video, or a podcast (4.5\%); and being told by someone else about a data breach (4\%). See Table~\ref{tab:threat-type-by-source-agg} for more detail. We find that different threat types are associated with different sources of information (${\chi}^2$: 496.85, p: < 0.001).

\subsubsection{Content Shared With Participants}
We also explored why and how participants thought that content had been shared with them. Of the 15\% of participants who had content shared directly with them, this was primarily received via a message (46\%) or face-to-face (31\%), vs social media (14\%), calls (1\%) or other means (8\%). Content primarily came from friends (32\%), colleagues (20\%), and other non-familial or workplace acquaintances (32\%). Information was thought to be shared because it was relevant to protect them (47\%), it described the sender's personal experience (21\%), it related to the respondents personal experience (19\%), it was of high profile (16\%), or the sender frequently shared (12\%). The most common combinations of sharer and threat type was news on a hacking incident that was shared from a friend (7.5\%) or by other means (8.6\%), followed by news on a data breach shared by a friend (7.5\%) or through other/social media channels (6.3\%) (${\chi}^2$: 87.104, p: 0.03, Table \ref{tab:who-shared-which-news}). Regression modelling showed age correlated with sharing rates, with older participants less likely have content shared with them (0.95, Table \ref{tab:regression-howdiscovered}).

\subsubsection{Sharing by Participants.}
Of the 26\% of participants who indicated they had shared their content with someone else, there was a more balanced distribution of recipients. Most common were partners (30.7\%), followed by family (26.1\%), colleagues (20.5\%), friends (20.1\%), and others (2.6\%). There were no significant effects of age or gender identity on whether participants reported actually sharing content with others. For comparison, in~\cite{10.1145/3173574.3173575} sharing was reported with friends (59\%), family (53\%), partners (34\%) and colleagues (19\%), with participants able to choose multiple options.

Privacy and security concerns were found to correlate with sharing behaviour, with those not sharing content with others having slightly lower levels of concern (IUIPC corr: 0.08, p:0.008, and SA6 corr: 0.16, p: < 0.0001). There was also a small correlation between lower privacy concern and content being shared with participants (IUIPC corr: 0.07, p:0.02). Whether respondents shared cybersecurity news with others was associated with threat type, with respondents less inclined to share phishing information (-6.4\%) and more inclined to share ransomware news (+5.6\%) or generic information on a data breach (+4.1\%). After correcting for both demographic factors and running multiple comparisons, there were no significant interactions between country of residence and sharing with or by participants.

\subsection{Effects Around Participation in Workplace Training (RQ2)}
\begin{figure*}[h]
\centering
\includegraphics[width = 0.6\textwidth]{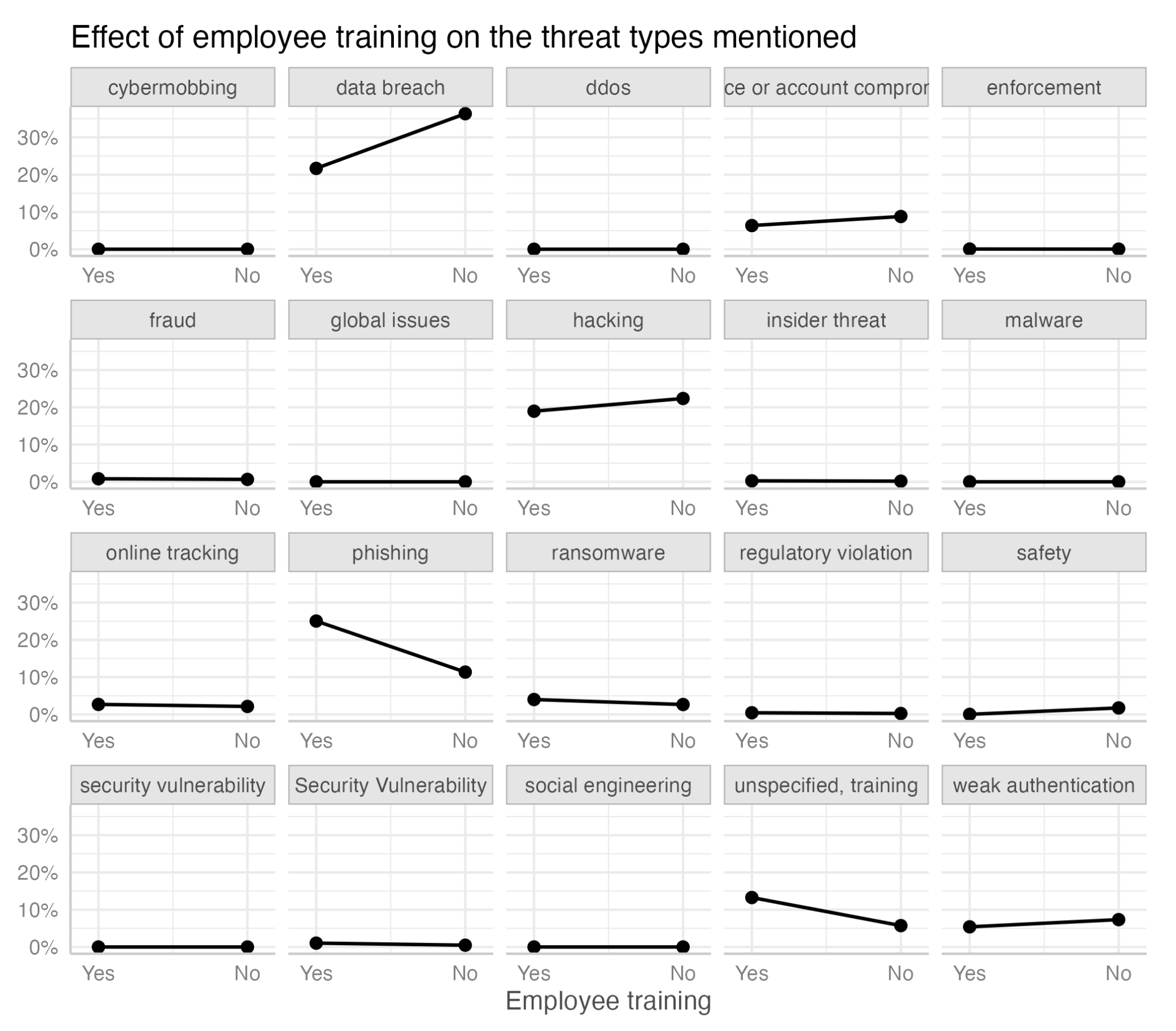}
\caption{Differences in content threat types between those who did and did not receive workplace training.}
\label{fig:content-category-by-training}
\end{figure*}

Participant engagement with workplace training differed considerably across the four countries in the study (Figure~\ref{fig:workplace-training-frequency}). While the study design could not prove that effects were caused by training, its often mandatory nature means that correlated changes are much more likely to have been caused by training than to have caused it. After accounting for demographic factors, those who had received workplace training were 27\% more likely to have described content provided by their employer (Table~\ref{tab:regression-howdiscovered}). Participation in workplace training also correlated with the types of threat that respondents mentioned (${\chi}^2$: 69.97, p: < 0.001), with people who had received training much more likely to have consumed content on phishing and those who had not more likely to refer to hacking, data breaches or weak authentication (Figure~\ref{fig:content-category-by-training}) and report lower baseline security concerns (SA6 corr: 0.24, p: <0.0001). Training participation was not correlated with actual onwards sharing (Table~\ref{tab:regression-whether-onward-shared}) but was associated with reduced intention to share content outside of work. The \textit{recipients} of cybersecurity content shared by participants who had undertaken workplace training was also different (${\chi}^2$: 105.98, p: < 0.001). Those with workplace training had a greater intention to share news with colleagues (+23.6\%) but were less likely to say they would share with those outside the professional environment such as friends (-8.9\%), family (-8.9\%), and spouses (-5.3\%). Correspondingly, participants were more likely to have had news shared with them by a colleague if they had received workplace training.

\section{Discussion}
\subsection{The Impact of Employers on Engagement With Cybersecurity}
The main impact observed was how the inclusion of employers as a source of cybersecurity content showed how many people's information comes from the workplace, going beyond a temporary jump in awareness following training. Given that the most common training frequency annual we estimate that only around 8\% of the participants receiving workplace training had done so in the month before the survey, yet employers were the top source of information across all participants for the last piece of cybersecurity information they had encountered. This is important as the dominance of phishing content from those who reported undertaking workplace training risks downplaying other threats faced by people and businesses. Phishing is a convenient training topic as it easily affords simulation exercises (compared to e.g., not writing passwords down), and focusing training on one attack vector simplifies the messaging that people receive. For training providers, it makes it easier to produce and demonstrate training effects to clients. It is, however, important that this does not result in the impression that phishing is the \textit{only} threat worth caring about.

\subsection{Corporate Protection at the Expense of Personal Protection}
The disparity in resources allocated to personal vs professional cybersecurity training risks current approaches communicating cybersecurity as a \textit{workplace} issue rather than a \textit{universal} one. While other kinds of workplace training like manual handling might give people the skills required to e.g., lift heavy objects at home, this analogy holds poorly for workplace cybersecurity training where there is a gap between business and personal risks. This could in turn cause problems for employers given the dramatic shift towards online and remote working seen since 2020, with personal and work devices increasingly overlapping~\cite{10.1145/2702123.2702386, 10.1145/3313831.3376881}. Compounded by the compliance focus of adjacent training (such as for the GDPR), employees and management may thus be inclined to consider themselves `trained' at work and not engage further at home. One way around this could be to offer licenses for security tools (e.g. password managers) to employees and their families. This practice could increase opportunities for employees to upskill their social environment and practice good security behaviour. The employer benefits from both.

\subsection{Effects Beyond the Workplace}\label{sec:effects-beyond-workplace}
The observed focus on phishing is likely to have effects beyond the workplace, as workplace training shapes what people consider to be `cybersecurity' and may be the only cybersecurity training that they receive. Notably, workplace training is usually less concerned with measures that are key at home but enforced by IT departments at work, such as choosing strong passwords, using multifactor authentication, and keeping devices up to date (all recommended by CISA). It is also no longer the case (if it ever was) that social engineering attacks utilise personal channels to target personal assets and workplace channels to target workplace assets. Turning cybersecurity into a workplace issue may therefore push attackers towards personal channels that employees are less cautious over and organisations cannot monitor. An opportunity exists to leverage the ability of workplace training programs to deliver opportunities for people to practice and apply their new skills. A consistent problem with public-oriented cybersecurity advice from governments/banks/etc. is that there is no accompanying opportunity to put things into practice; your bank might have told you about the key elements of a scam, but without the opportunity to see a fraudulent message in context. If the scope of workplace training was widened to include these `personal' scenarios, then this would likely improve people's cybersecurity awareness in both personal and business contexts. This could make training more relevant, with the key motivations for sharing seen in the survey being protecting others and sharing personal experiences.

\begin{acks}
This work was supported by KnowBe4.
\end{acks}

\bibliographystyle{ACM-Reference-Format}
\bibliography{main}

\clearpage

\section{APPENDIX}
\begin{table*}[!htbp] \centering 
\begin{tabular}{@{\extracolsep{5pt}}lccc} 
 & \multicolumn{3}{c}{\textit{Dependent variable:}} \\ 
\cline{2-4} 
\\[-1.8ex] & phone\_or\_other & social & text \\ 
\\[-1.8ex] & (1) & (2) & (3)\\ 
\hline \\[-1.8ex] 
 age & 1.013 & 0.978 & 1.041$^{**}$ \\ 
  & (0.031) & (0.025) & (0.018) \\ 
  & p = 0.678 & p = 0.369 & p = 0.021 \\ 
  & & & \\ 
 genderMale & 0.157$^{*}$ & 1.355 & 1.229 \\ 
  & (1.102) & (0.516) & (0.403) \\ 
  & p = 0.094 & p = 0.557 & p = 0.610 \\ 
  & & & \\ 
 employeeTrainingYes & 1.064 & 1.129 & 2.339$^{**}$ \\ 
  & (0.635) & (0.483) & (0.382) \\ 
  & p = 0.923 & p = 0.802 & p = 0.027 \\ 
  & & & \\ 
 Constant & 0.222 & 0.910 & 0.194$^{**}$ \\ 
  & (1.051) & (0.806) & (0.648) \\ 
  & p = 0.153 & p = 0.907 & p = 0.012 \\ 
  & & & \\ 
\hline \\[-1.8ex] 
Akaike Inf. Crit. & 406.490 & 406.490 & 406.490 \\ 
\end{tabular} 
  \caption{How content was shared with participants. Items in bold were found to be significant after correction for multiple comparisons.} 
  \label{tab:regression-how-inward-shared} 
\end{table*}

\begin{table*}[!htbp] \centering 
\begin{tabular}{@{\extracolsep{5pt}}lcccccc} 
  & \multicolumn{6}{c}{\shortstack{\textit{Dependent variable:}\\\textbf{}\\constant: email or message from 3rd party}} \\ 
\cline{2-7} 
\\[-1.8ex] & employer  & Other & shared & social media & TV, radio & website \\ 
\\[-1.8ex] & (1) & (2) & (3) & (4) & (5) & (6)\\ 
\hline \\[-1.8ex] 
 age & 0.962 & 0.988 & \textbf{0.950} & 0.956 & 0.998 & 0.967 \\ 
  & (0.015) & (0.017) & \textbf{(0.015)} & (0.015) & (0.015) & (0.015) \\ 
  & p = 0.010 & p = 0.482 & \textbf{p = 0.001} & p = 0.004 & p = 0.893 & p = 0.026 \\ 
  & & & & & & \\ 
 genderMale & 0.251 & 0.924 & 0.091 & 1.068 & 0.710 & 0.601 \\ 
  & (0.848) & (1.059) & (0.864) & (0.854) & (0.882) & (0.806) \\ 
  & p = 0.104 & p = 0.941 & p = 0.006 & p = 0.939 & p = 0.698 & p = 0.528 \\ 
  & & & & & & \\ 
 employeeTrainingYes & \textbf{11.017} & 0.632 & 1.503 & 0.826 & 0.743 & 0.986 \\ 
  & \textbf{(0.329)} & (0.354) & (0.279) & (0.277) & (0.289) & (0.263) \\ 
  & \textbf{p = 0.000} & p = 0.196 & p = 0.145 & p = 0.490 & p = 0.304 & p = 0.959 \\ 
  & & & & & & \\ 
 countryofresidenceGermany & 1.554 & 0.777 & 0.671 & 0.398 & 0.425 & 0.573 \\ 
  & (0.419) & (0.464) & (0.412) & (0.407) & (0.464) & (0.381) \\ 
  & p = 0.293 & p = 0.587 & p = 0.333 & p = 0.024 & p = 0.065 & p = 0.145 \\ 
  & & & & & & \\ 
 countryofresidenceUnited Kingdom & 1.649 & 0.565 & 0.851 & 0.582 & 1.194 & 0.469 \\ 
  & (0.422) & (0.513) & (0.421) & (0.417) & (0.428) & (0.400) \\ 
  & p = 0.237 & p = 0.266 & p = 0.703 & p = 0.195 & p = 0.680 & p = 0.059 \\ 
  & & & & & & \\ 
 countryofresidenceUnited States & 0.972 & 0.372 & 0.997 & 0.879 & 0.895 & 0.548 \\ 
  & (0.428) & (0.526) & (0.407) & (0.392) & (0.424) & (0.385) \\ 
  & p = 0.948 & p = 0.060 & p = 0.995 & p = 0.743 & p = 0.793 & p = 0.119 \\ 
  & & & & & & \\ 
 age:genderMale & 1.027 & 0.993 & \textbf{1.051} & 0.999 & 1.010 & 1.029 \\ 
  & (0.021) & (0.027) & (0.021) & (0.022) & (0.021) & (0.020) \\ 
  & p = 0.202 & p = 0.782 & p = 0.019 & p = 0.967 & p = 0.633 & p = 0.153 \\ 
  & & & & & & \\ 
 Constant & 1.776 & 2.599 & 15.309 & 16.739 & 2.131 & 11.419 \\ 
  & (0.650) & (0.713) & (0.612) & (0.618) & (0.651) & (0.611) \\ 
  & p = 0.378 & p = 0.181 & p = 0.00001 & p = 0.00001 & p = 0.246 & p = 0.0001 \\ 
  & & & & & & \\ 
\hline \\[-1.8ex] 
Akaike Inf. Crit. & 3,848.531 & 3,848.531 & 3,848.531 & 3,848.531 & 3,848.531 & 3,848.531 \\ 
\end{tabular}
  \caption{How participants discovered content. Items in bold were found to be significant after correction for multiple comparisons.} 
  \label{tab:regression-howdiscovered} 
\end{table*} 

\begin{table*} \centering 
\begin{tabular}{@{\extracolsep{5pt}}lc} 
\\[-1.8ex]\hline 
\hline \\[-1.8ex] 
 & \multicolumn{1}{c}{\textit{Dependent variable:}} \\ 
\cline{2-2} 
\\[-1.8ex] & sharedOnwardYes \\ 
\hline \\[-1.8ex] 
 age & 0.990 \\ 
  & (0.008) \\ 
  & p = 0.212 \\ 
  & \\
 genderMale & 1.012 \\ 
  & (0.434) \\ 
  & p = 0.978 \\ 
  & \\ 
 Q14Yes & 0.873 \\ 
  & (0.146) \\ 
  & p = 0.351 \\ 
  & \\ 
 countryofresidenceGermany & 0.883 \\ 
  & (0.203) \\ 
  & p = 0.542 \\ 
  & \\ 
 countryofresidenceUnited Kingdom & 1.344 \\ 
  & (0.214) \\ 
  & p = 0.166 \\ 
  & \\ 
 countryofresidenceUnited States & 0.999 \\ 
  & (0.204) \\ 
  & p = 0.998 \\ 
  & \\ 
 age:genderMale & 0.995 \\ 
  & (0.011) \\ 
  & p = 0.641 \\ 
  & \\ 
 Constant & 4.760$^{***}$ \\ 
  & (0.320) \\ 
  & p = 0.00001 \\ 
  & \\ 
\hline \\[-1.8ex] 
Observations & 1,092 \\ 
Log Likelihood & $-$618.930 \\ 
Akaike Inf. Crit. & 1,253.861 \\ 
\hline 
\hline \\[-1.8ex] 
\textit{Note:}  & \multicolumn{1}{r}{$^{*}$p$<$0.1; $^{**}$p$<$0.05; $^{***}$p$<$0.01} \\ 
\end{tabular} 
  \caption{Whether participants shared their content with anyone else. Items in bold were found to be significant after correction for multiple comparisons.} 
  \label{tab:regression-whether-onward-shared} 
\end{table*} 

\begin{table*}
    \centering
    \begin{tabularx}{\linewidth}{r|Xr|Xr|Xr|Xr}
        & Website & & Broad/Podcast & & Shared & & Social Media & \\
        \hline
        1 & Data Breach & 5.5\% & Data Breach & 4.5\% & Data Breach & 4\% & Hacking & 3.7\% \\ 
        2 & Hacking & 3.4\% & Hacking & 2.3\% & Hacking & 3\% & Data Breach & 3\% \\ 
        3 & General Training & 2.5\% & Ransomware & 1.1\% & Phishing & 2.5\% & Phishing & 1.6\% \\ 
        \hline
        & Company & & Employer & & Other & & & \\
        \hline
        1 & Data Breach & 2.6\% & Phishing & 7.5\% & Data Breach & 1.2\% & & \\
        2 & Phishing & 1.6\% & General Training & 5.3\% & Phishing & 1\% & & \\
        =3 & Weak Auth. & 1.2\% & Data Breach & 1.7\% & General Training & 0.7\% & & \\
        =3 & - & & - & & Weak Auth. & 0.7\% & & \\
    \end{tabularx}
    \caption{The most common threat types received through different sources.}
    \label{tab:threat-type-by-source-agg}
\end{table*}

\begin{table*}
    \centering
    \begin{tabular}{r|l|r|l|r|l|r}
        & Colleague & & Family Member & & Friend & \\
        \hline
        1 & Data Breach & 4.6\% & Data Breach & 2\% & Hacking & 7.5\% \\
        2 & Phishing & 4\% & Hacking & 2\% & Data Breach & 6.6\% \\
        =3 & Hacking & 2.6\% & Phishing & 2\% & Weak Authentication & 3.7\% \\
        =3 & Weak Authentication & 2.6\% & & & & \\
        \hline
         & Spouse/Partner & & Other & & & \\
        \hline
        1 & Data Breach & 2.6\% & Hacking & 8.6\% & & \\
        2 & Phishing & 1.4\% & Data Breach & 6.3\% & & \\
        =3 & Device or Account Compromise & 0.6\% & Phishing & 2.9\% & & \\
        =3 & Weak Authentication & 0.6\% & & & & \\
    \end{tabular}
    \caption{The most common content types shared through different relationships.}
    \label{tab:who-shared-which-news}
\end{table*}

\begin{table*}
\centering
\footnotesize
\setlength\tabcolsep{5pt}
\begin{tabularx}{\linewidth}{XrXrXrXrXr}
   \hline
 \zz{2}{A colleague} & \zz{2}{A family member } & \zz{2}{A friend} & \zz{2}{My spouse/partner}& \zz{2}{Other} \\ 
  \hline
data breach & 4.6\% & data breach & 2\% & hacking & 7.5\% & data breach & 2.6\% & hacking & 8.6\% \\ 
  phishing & 4\% & hacking & 2\% & data breach & 6.6\% & phishing & 1.4\% & data breach & 6.3\% \\ 
  hacking & 2.6\% & phishing & 2\% & weak authentication & 3.7\% & device or account compromise & 0.6\% & phishing & 2.9\% \\ 
  weak authentication & 2.6\% & weak authentication & 1.1\% & device or account compromise & 2.9\% & weak authentication & 0.6\% & weak authentication & 2.3\% \\ 
  unspecified, training & 2.3\% & online tracking & 0.9\% & phishing & 2.3\% & hacking & 0.3\% & device or account compromise & 2\% \\ 
  Security Vulnerability & 1.4\% & ransomware & 0.9\% & online tracking & 2\% & unspecified, training & 0.3\% & unspecified, training & 2\% \\ 
  global issues & 0.6\% & regulatory violation & 0.3\% & unspecified, training & 2\% & ddos & 0\% & Security Vulnerability & 1.7\% \\ 
  online tracking & 0.6\% & unspecified, training & 0.3\% & global issues & 1.7\% & fraud & 0\% & online tracking & 1.4\% \\ 
  regulatory violation & 0.6\% & ddos & 0\% & ransomware & 1.4\% & global issues & 0\% & fraud & 1.1\% \\ 
  device or account compromise & 0.3\% & device or account compromise & 0\% & fraud & 1.1\% & malware & 0\% & ransomware & 1.1\% \\ 
  fraud & 0.3\% & fraud & 0\% & ddos & 0.3\% & online tracking & 0\% & regulatory violation & 1.1\% \\ 
  malware & 0.3\% & global issues & 0\% & safety & 0.3\% & ransomware & 0\% & global issues & 0.6\% \\ 
  social engineering & 0.3\% & malware & 0\% & Security Vulnerability & 0.3\% & regulatory violation & 0\% & malware & 0.3\% \\ 
  ddos & 0\% & safety & 0\% & social engineering & 0.3\% & safety & 0\% & safety & 0.3\% \\ 
  ransomware & 0\% & Security Vulnerability & 0\% & malware & 0\% & Security Vulnerability & 0\% & ddos & 0\% \\ 
  safety & 0\% & social engineering & 0\% & regulatory violation & 0\% & social engineering & 0\% & social engineering & 0\% \\ 
   \hline
\end{tabularx}
\label{tab:who-shared-which-news-detail}
   \caption{Respondents received different cybersecurity content through different relationships (${\chi}^2$: 87.104, p: 0.03).}
\end{table*}
\end{document}